\documentclass{emulateapj}
\usepackage{times}

\def\la{\mathrel{\mathpalette\fun <}}
\def\ga{\mathrel{\mathpalette\fun >}}

\def\simpropto{\lower.2ex\hbox{$\; \buildrel \sim \over \propto \;$}}

\def\fun#1#2{\lower0.837ex\vbox{\baselineskip0ex\lineskip0.209ex
  \ialign{$\mathsurround=0ex#1\hfil##\hfil$\crcr#2\crcr\sim\crcr}}}

\def\msun{M_\odot}

\def\sles{\lower2pt\hbox{$\buildrel {\scriptstyle <}
   \over {\scriptstyle\sim}$}}

\def\sgreat{\lower2pt\hbox{$\buildrel {\scriptstyle >}
   \over {\scriptstyle\sim}$}}

\def\la{\mathrel{\mathpalette\fun <}}
\def\ga{\mathrel{\mathpalette\fun >}}

\def\msun{M_\odot}

\begin{document}

\title{The Shape of Long Outbursts in U Gem type
    Dwarf Novae from AAVSO data}
\shortauthors{CANNIZZO}
\author{
 John~K.~Cannizzo\altaffilmark{1,2}
   }
\altaffiltext{1}{CRESST and Astroparticle Physics Laboratory
               NASA/GSFC, Greenbelt, MD 20771, USA
               John.K.Cannizzo@nasa.gov}
\altaffiltext{2}{Department of Physics, University of Maryland,
              Baltimore County, 1000 Hilltop Circle,
              Baltimore, MD 21250, USA}

\begin{abstract}
We search the American Association of Variable Star Observers 
   (AAVSO) archives of the two best studied
  dwarf novae in an attempt to find light curves for
   long outbursts that are extremely well-characterized.
The systems are U Gem and SS Cyg.
   Our goal is to search for embedded precursors
  such as those  that have  been found
   recently in the high fidelity
   {\it Kepler}  data for superoutbursts
  of some members of the SU UMa subclass of dwarf novae. 
        For the vast majority of AAVSO data, the combination
  of low data cadence and large errors associated with individual
   measurements precludes one from making any strong statement
 about the shape of the long outbursts.
   However, for a small number of outbursts, extensive
long term monitoring with digital photometry yields high fidelity 
  light curves.  We report the discovery of embedded 
 precursors in two of three candidate long
   outbursts.  
   {\it This is the first time that such embedded precursors  
 have been found in dwarf novae above the period gap,} and
       reinforces van Paradijs' finding 
    that long
outbursts in dwarf novae above the period gap and superoutbursts in systems
below the period gap constitute a unified class.
   The thermal-tidal instability to account for superoutbursts
  in the SU UMa stars predicts embedded precursors only for short
  orbital period dwarf novae, therefore the presence  of embedded precursors
  in long orbital period systems $-$  U Gem and SS Cyg $-$
argues for a more general mechanism to explain long outbursts.
\end{abstract}

\keywords{ accretion, accretion disks --- binaries:
   close --- binaries: general --- novae, cataclysmic variables
  --- methods: observational
  --- stars: dwarf novae}

\section{Introduction}

%
{\it Kepler} observations 
  of short orbital period
   dwarf novae (below the ``period gap'')
   have given us
  for the first time clear evidence for the presence of
embedded precursors, or ``failed'' 
   outbursts,
    at the beginning of  superoutbursts of SU UMa
systems  
(Cannizzo et al. 2012).
  This behavior had 
  been partially seen before in
some systems with fragmentary data, but is now quite clear. 
   Thanks to the high fidelity AAVSO
data of the past $\sim$10 yrs,
   we report the discovery of such precursors also in two systems
{\it above} the period gap.
   This may have consequences for theoretical models for the
long outbursts and the superoutbursts.

Cataclysmic variables (CVs $-$ Warner 1995ab) are
  semi-detached interacting binaries 
  containing  a
  Roche-lobe filling K or M secondary that transfers 
          matter to a white dwarf (WD).
  CVs 
    show a ``gap''
between $P_h\sim$$2$ and 3
 (where $P_h = P_{\rm orbital}/1$ hr)
    during which
  time the 
  secondary star loses contact with its Roche lobe
  and mass transfer ceases as the systems evolve to shorter orbital periods.
  Thus at $P_h\simeq3$ the binary becomes fully detached.
       At $P_h\simeq2$ the secondary  comes back into 
contact with its Roche lobe and mass transfer resumes.
    Systems can also be ``born'' in the gap, so it is not
completely empty.
   For $P_h \la 2$ angular momentum loss
   from the binary is
   thought to be due solely to gravitational radiation.
The CV subclass of dwarf novae (DNe) 
   also have
   semi-periodic 
   outbursts. 
   SU UMa stars are DNe below the period gap
   exhibiting  short, normal outbursts (NOs) and superoutbursts (SOs).
   SOs show superhumps which are modulations in the light curve
   at periods slightly exceeding the orbital  period;
   superhumps are the defining property of SOs.

   DNe outbursts are thought to be due to a 
 thermal limit cycle accretion disk
  instability (Smak 1984)
   in which material is accumulated in quiescence
  and then dumped onto the WD during outburst.
 During short outbursts
   in longer period DNe,
  a  few percent of the stored mass is accreted, and during long outbursts
   a significant fraction $\sim$0.2 of the stored mass is accreted.
         For the SU UMa stars, a SO  is thought to accrete   $\ga0.7-0.8$
  of the stored mass.  
  Although the accretion disk is never in steady state 
   during the limit cycle, it is close 
   to steady state during 
long outbursts.

U Gem and SS Cyg are the two DNe
   with the most complete 
   AAVSO coverage.
U Gem ($P_h=4.25$) 
  was discovered by John Russell Hind in 1855 (Hind 1856),
  and 
 SS Cyg ($P_h=6.60$) 
  by Louisa D. Wells in 1896 (Wells 1896).
   To date, U Gem has $\ga$115,000
 observations, and SS Cyg has $\ga$455,000.
   Figure 1   
   shows the number of individual observations
  $n_{\rm obs}$  for each 24 hr interval versus
   time for each long term light curve.

\begin{figure}
\centering
\epsscale{1.0}
\includegraphics[scale=0.45]{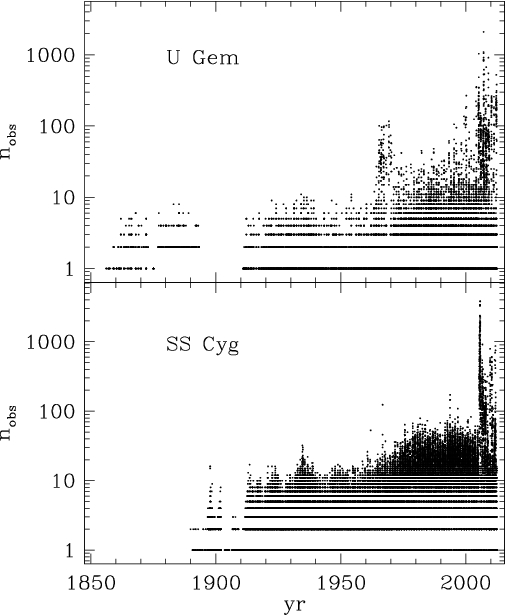}
\vskip -0.05cm
\figcaption{ 
   The 
   number of observations entering
  into a daily mean $n_{\rm obs}$
   for the historical AAVSO light
curves of U Gem ({\it top panel})
   and  SS Cyg  ({\it bottom panel}).
   }
\smallskip
\end{figure}

Recent observations by {\it Kepler}
have revealed details of the outburst behavior
 that have been partially seen previously\footnote{For
  example,
Figure 3.36 from Warner (1995b) shows precursors
 of varying depths in SOs of VW Hyi,
  but the light curves are smoothed, filled
  versions based on fragmentary data.},
     but which can now be
studied in much  greater detail 
  (Cannizzo et al. 2012).
   Of particular interest for this
  work is the presence  
   of an embedded ``failed'' NO
at the start of  a
  SO.
   Within the context of the thermal-tidal instability (TTI)
model for SOs (Osaki 1989;
       Ichikawa \& Osaki 1992; 
   Osaki 2005, see his Figures 3 and 4), 
  embedded precursors
   are understood
  as being due to a temporary
squeezing of the outer accretion disk
  by increased effective tidal forces
  due to the onset of a tidal
  instability.  This  instability
 drives the outer disk
 between circular and eccentric shapes
  when the outer edge of the disk 
expands beyond the point of 3:1  
  resonance with the binary orbital period,
i.e., the point at which $2\pi/\Omega$ 
   around 
the WD equals $P_{\rm orbital}/3$ (Whitehurst 1988).
   The discovery of the tidal instability  
 by Whitehurst
  led Osaki to propose the 
    TTI, 
     which combines the
      accretion disk
       thermal limit cycle instability 
   with the tidal instability.
   A necessary condition for the tidal instability
   is that the mass ratio $q\equiv M_2/M_1 < 0.25$
  so that for long outbursts, in which a substantial amount of
  matter is stored in the disk, the presence of high
viscosity material can expand the outer disk radius
  beyond the 3:1 radius. Thus superhumps are seen only
  in short orbital period DNe.

An important point is that, if the TTI model 
  only applies to low $q$ binaries, and therefore only
to short orbital period DNe, one expects it not 
 to apply to systems longward of the
period gap
   (since
for Roche-lobe filling main-sequence donors,
   systems with $q \la 0.25$ have $P_h \la 2.5$ hr).
  Therefore, the presence
of embedded precursors in systems above the period gap,
   should they exist,
would argue for a more general physical mechanism 
     for long outbursts in all DNe.
Previous studies of
  DNe outbursts using amateur data have
   proved useful
   in delineating
  timescales
   and constraining models
(e.g., 
          Campbell  \& Shapley 1940,
 Sterne,  Campbell  \& Shapley 1940,
          Bath \& van Paradijs 1983,
                  van Paradijs 1983,
             Szkody  \& Mattei 1984,
            Cannizzo \& Mattei 1992, 
                               1998,
                    Ak, et al. 2002,
         Simon 2004).
Van Paradijs 
    (1983)
  studied 
 a sample of  
     DNe 
  spanning the
period gap
  and
  found that 
 short outburst
   durations
   increase with orbital period,
   whereas 
long outburst
  durations  
   are relatively constant 
  with orbital period.
 He
   proposed that the
     relation of SOs to NOs
for DNe below the period gap and
  of  
   long outbursts to NOs 
  for DNe above the period gap are equivalent; 
  superoutbursts are, in his view,
     just long outbursts
  seen in 
   short orbital period DNe.
   This finding was amplified by Ak et al. (2002)
using a larger sample.

For completeness we note that there have been 
 prior 
critical analyses of the TTI model, 
   from both theoretical and observational perspectives.
 Schreiber et al. (2004)
 compare numerical time dependent
   models of the TTI
   with an alternative model $-$
  enhanced mass transfer (EMT) from the secondary star.
   Their arguments are somewhat general, however,
  and although they conclude in the end that
the EMT is favored, they note ``we have not proven
  the EMT [model] to be correct nor the TTI [model]
   not to work.''
   Also, in a series of papers Smak (2009abcd)
        presents  arguments against the standard
  interpretation of superhumps in the SU UMa stars
as being due to a precessing, eccentric disk\footnote{The
most recent numerical simulations of 
  an accretion disk
subject to the 3:1 instability  reveal a disk shape that
alternates between eccentric and circular (Montgomery
   2012ab), 
  therefore it may be more difficult than
previously thought to detect an (instantaneously) 
  eccentric disk shape in 
  an eclipsing system.}.
  Thus from his perspective 
the entire physical basis
  for the TTI would be called into question. 
  Smak also favors the EMT model.
  Cannizzo et al. (2012)
 study   the long term {\it Kepler} light curves
  of two SU UMa stars  and argue  that the 
tendency of 
recurrence times for normal outbursts 
to exhibit sometimes a local
maximum half way between   two superoutbursts
  argues against the TTI model,
  wherein one expects a monotonically increasing
  series of recurrence times  between  two superoutbursts
  (Ichikawa, Hirose, \& Osaki 1993, see their Figure 1;
                        Osaki 2005,  see  his Figure 3).

\section{High Fidelity Long Outbursts in U Gem and SS Cyg}

Figure 1  
    shows the number of
   individual observations $n_{\rm obs}$
   for each 24 hr interval within the historical
   AAVSO light curves for U Gem and SS Cyg.
   Data prior to 2000 are mainly visual and the errors are
   too large to be usable for
 detecting subtle effects such as embedded precursors to outbursts.

   One can see a period of
 faster-than-exponential growth in
  the upper envelope of $n_{\rm obs}$ versus time.
   For the vast majority of the data, 
   the  uncertainty in flux
  associated with an individual measurement
     produces a scatter in  the
  light curve
     precludes any detailed study of the shape of 
the outburst.   Historically, the majority of the
   data were from visual observations
   in which estimates of a given variable
 star brightness are carried out 
           by comparison with nearby field stars.
    The precision for any given data point in this method
   is $\sim$0.3-0.5 mag, and, 
given the varying color biases between 
   different observers,
   a simple averaging 
 of $m_V(t)$ values for outburst during   
     times when $n_{\rm obs}$
  is large does not guarantee a light curve
   with true physical fidelity.

   In the last $\sim$10 yr this situation has
changed due to the increased use of digital photometry.
   This trend can be seen in Figure 1.  
  The intervals with digital photometry
generally are accompanied by times during which $n_{\rm obs} \ga$$300-1000$,
i.e., more than $\sim$$300-1000$ observations
   within a 24 hr interval.
  By restricting our attention to long outbursts
lying within these times of large $n_{\rm obs}$
  during the last 10 yr,
  and furthermore restricting the data to digital
 $V-$band photometry only (so as exclude the individual color 
biases inherent in the visual data), we may
  considerably reduce the vertical scatter in the outburst
data.
  We note that the presence
    of high $n_{\rm obs}$
  data is a necessary but not a sufficient ingredient.
  High  $n_{\rm obs}$ values  ($\ga300$) 
  are indicative of
        the use of digital photometry, 
   but there must also be a high cadence rate near outburst 
  onset to 
   have  a reasonable chance
   of  determining whether
  or not a failed, embedded outburst
   accompanies the outburst start.

%

\begin{figure}
\centering
\epsscale{1.0}
\includegraphics[scale=0.45]{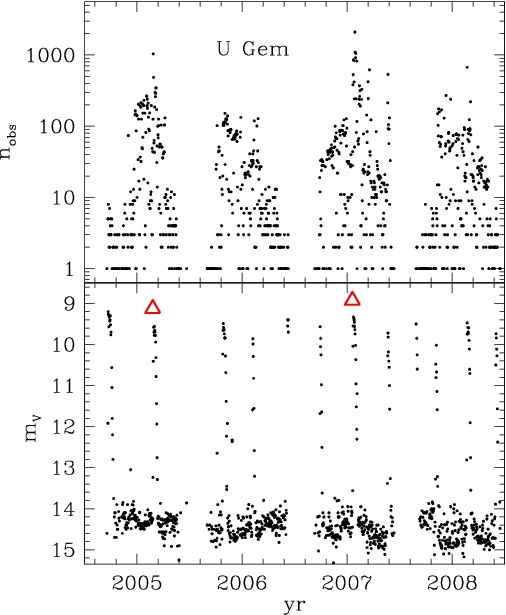}
\vskip -0.05cm
\figcaption{ 
  A portion of the U Gem 
  long term light curve
   containing
   points for which $n_{\rm obs}$ is large. 
 Shown are $n_{\rm obs}$ ({\it top panel})
  and daily means for $m_V$ ({\it bottom panel}). 
   The small triangles ({\it red}) indicate the 
high fidelity long outbursts.
   }
\smallskip
\end{figure}

\begin{figure}
\centering
\epsscale{1.0}
\includegraphics[scale=0.45]{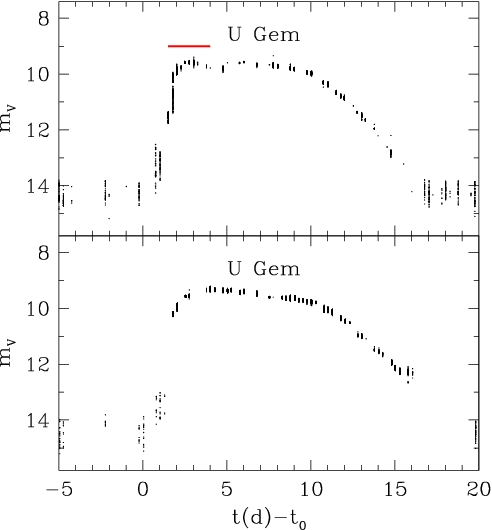}
\vskip -0.05cm
\figcaption{
Detailed views for U Gem
   showing the first
  outburst of 2005 ({\it top panel}), for which $t_0=2005.15$,
   and the first
  outburst of 2007 ({\it bottom panel}), for which $t_0=2007.05$.
 Here we plot individual measurements
  rather than daily means.
       Only
   digital $V-$band photometry data are plotted.
   The horizontal bar ({\it red}) indicates
    the embedded precursor.
       }
\smallskip
\end{figure}

We have made careful study of each long outburst
in U Gem and SS Cyg since 2000  in an effort to identify
  outbursts that are sufficiently well sampled by digital 
photometric observations such that one may make a clear statement
concerning the presence of an embedded precursor.
  There are two such outbursts in U Gem and one in SS Cyg.
Figure 2 shows an enhanced view of the 
  U Gem 
  light curve 
   containing the two candidate
  long outbursts.
   The first outbursts of 2005 and 2007
are long and  lie with regions of dense digital
 photometric coverage.
  Figure 3 shows detailed views of these two outbursts.
 The 2005 outburst shows an apparent failed outburst at the start,
  and the 2007 outburst does not.
  Figure 4 shows a 
 stretch of  $n_{\rm obs}>1000$
   data for SS Cyg containing our one identified high fidelity
  long outburst, 
   at  $\sim$2005.7.
   It is shown in detail in Figure 5.
  It appears also to contain a failed outburst
at the start.
   In summary, for two of the three long outbursts
 in the best studied DNe 
  (both of which  lie above the period gap)
   for which any
 statement can be made about the detailed 
  shape of the outburst,
   there is an 
   apparent
failed, embedded outburst at the 
start, followed by a period  of slow decay,
   followed by a period of faster decay.
  These three characteristics of long
   outbursts are the same as in the superoutbursts of 
 the SU UMa stars,
  which lie below the period gap.

\begin{figure}
\centering
\epsscale{1.0}
\includegraphics[scale=0.45]{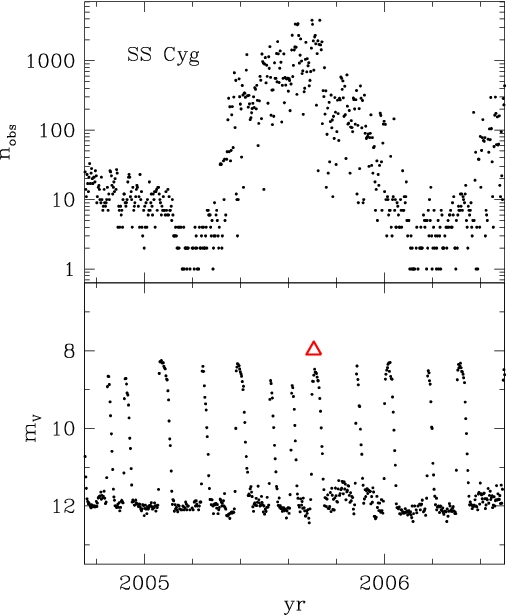}
\vskip -0.05cm
\figcaption{
 A detailed view for SS Cyg of the portion
  of the long term light curve containing
   points for which $n_{\rm obs}$ is large.
     Shown are $n_{\rm obs}$ ({\it top panel})
  and daily means for $m_V$ ({\it bottom panel}).
   The small triangle ({\it red}) indicates the
high fidelity long outburst.
             }
\smallskip
\end{figure}

\begin{figure}
\centering
\epsscale{1.0}
\includegraphics[scale=0.45]{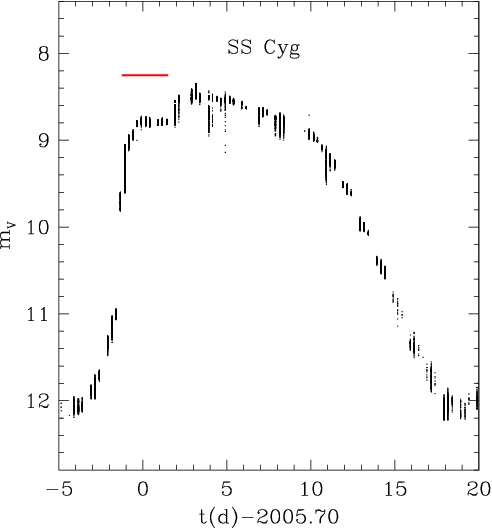}
\vskip -0.05cm
\figcaption{
  A detailed view of the
  high fidelity long outburst
   during late 2005 in SS Cyg, 
   showing only
   digital $V-$band photometry.
    Individual measurements are shown,
 not daily means. 
   The horizontal bar ({\it red}) indicates
    the embedded precursor.
       }
\smallskip
\end{figure}

\section{Discussion and Conclusion}
 
By examining the best AAVSO data for U Gem and SS Cyg
we have found evidence for embedded failed outbursts
  at the start of two out of three long outbursts
for which such an exercise is feasible.
   These data are from digital photometry.
For the 2005 U Gem outburst, the possibility of 
  a time variable calibration to explain the embedded 
precursor seems unlikely since the light curve is
   comprised of data from several observers.
For the 2005 SS Cyg outburst, the precursor stands out
more strongly.
   For other data in the long term light curves,
    the quality is not good enough  for one
 to be able to make any clear statement about the detailed
outburst shape.

The thermal-tidal instability model was developed 
to account for the short orbital period SU UMa stars
which have $q < 0.25$. 
   Since CVs have Roche-lobe  filling secondary 
 stars, this constrains
    the secondary star mass 
  to be 
  $\sim$$0.1\msun P_h$,
   unless the star has been 
   driven considerably out of thermal
  equilibrium due to mass loss.
 Therefore DNe above the period gap such
  as U Gem and SS Cyg with orbital periods 
   between 4 and 7 hr cannot possibly satisfy $q < 0.25$
  for $M_{\rm WD}\simeq \msun$,
   and therefore the thermal-tidal model 
   should not be a physical ingredient
 of  their outbursts.
   Two of three
   long outbursts in DNe longward of the period
   gap which have the requisite AAVSO
 data fidelity 
   show 
    (i) the initial failed outburst embedded at the start,
   (ii) the slow decay consistent with viscous decay, and
  (iii) the faster decay consistent with thermal decay.
These three characteristics
  are also manifested
  in the  long outbursts in DNe below the period gap
$-$  i.e., the superoutbursts in the SU UMa stars.
   Therefore Occam's razor would seem to demand a common 
   explanation which does not depend on $q$.
 The only difference between the two types of outbursts
is the presence of superhumps in the superoutbursts, 
   thus the association of  superhumps with  superoutbursts
 appears  to be associative rather than causal.

Finally, it is worth noting that U Gem exhibited an unusually
long outburst in October 1985 with a duration
of $\sim$39 d (Cannizzo et al. 2002). The usual duration
of  long outbursts in U Gem is $\sim$12 d.
   It is unclear how the existence of this long outburst 
   impacts our conclusions.
In the context of the standard accretion disk
limit cycle for DN outbursts (Smak 1984),
 the cooling front responsible for terminating the flat-topped
portion of a long outburst must have been prevented 
from propagating, perhaps due to an unusually 
  large amount of stored disk
mass.  
Cannizzo et al. (2002)
  determined a viscous decay time $26\pm6$ d mag$^{-1}$
from the AAVSO light curve.
   Given that virtually all other long outburst  decays
  in U Gem and SS Cyg do not last long enough for the decay
slope to be measurable, this one long outburst  gives
us our only measure of the viscous time scale in a DN 
 longward of the period gap.

\smallskip

We acknowledge the dedication and perseverance of the thousands 
 of observers contributing data to the AAVSO International Database.
 We thank Allen Shafter for useful comments.

\def\mnras{MNRAS}
\def\apj{ApJ}
\def\apjs{ApJS}
\def\apjl{ApJL}
\def\aj{AJ}
\def\araa{ARA\&A}
\def\aap{A\&A}
\def\aapl{A\&AL}
\def\pasj{PASJ}

\vfil\eject

\end{document}